\documentclass[useAMS,usenatbib]{mn2e}

\usepackage[dvips]{graphicx}
\usepackage{amssymb}
\usepackage{amsmath}
\usepackage{amstext}

\newcommand{\te}{$T_{e}$}
\newcommand{\nel}{$n_{e}$}

\newcommand{\ha}{H$\alpha$}
\newcommand{\hb}{H$\beta$}
\newcommand{\hd}{H$\delta$}
\newcommand{\hg}{H$\gamma$}
\newcommand{\hei}{He~{\sc i}\relax}
\newcommand{\nii}{[N~{\sc ii}\relax]}
\newcommand{\sii}{[S~{\sc ii}\relax]}
\newcommand{\siii}{[S~{\sc iii}\relax]}
\newcommand{\oi}{[O~{\sc i}\relax]}
\newcommand{\oii}{[O~{\sc ii}\relax]}
\newcommand{\oiii}{[O~{\sc iii}\relax]}
\newcommand{\feii}{[Fe~{\sc ii}\relax]}
\newcommand{\feiii}{[Fe~{\sc iii}\relax]}
\newcommand{\neiii}{[Ne~{\sc iii}\relax]}
\newcommand{\Hp}{H$^{+}$\relax} 
\newcommand{\Oo}{O$^{0}$\relax} 
\newcommand{\Op}{O$^{+}$\relax} 
\newcommand{\Opp}{O$^{2+}$\relax} 
\newcommand{\Np}{N$^{+}$\relax} 
\newcommand{\Sp}{S$^{+}$\relax} 
\newcommand{\Hep}{He$^{+}$\relax} 

\newcommand\ion[2]{#1~{\sc {#2}}\relax}        
\newcommand\ioni[2]{${\rm #1^{#2}}$}           

\newcommand{\cmc}{{\rm cm$^{-3}$}}
\newcommand{\kms}{{\rm km~s$^{-1}$}}

\newcommand{\hhu}{HH~203}
\newcommand{\hhd}{HH~204}
\newcommand{\oric}{$\theta^1$~Ori~C}
\newcommand{\oria}{$\theta^2$~Ori~A}
\newcommand{\hii}{H~{\sc ii}}
\newcommand{\hi}{H~{\sc i}}
\newcommand{\ld}{$\lambda$}

\title[Bidimensional spectroscopy of \hhd]{Exploring the effects of high-velocity flows in abundance 
       determinations in \hii\ regions. Bidimensional spectroscopy of \hhd\ in the Orion Nebula%
			       \thanks{Based on observations collected at the Centro Astron\'omico 
			       Hispano Alem\'an (CAHA) at Calar Alto, operated jointly by the 
			       Max-Planck Institut f\"ur Astronomie and the Instituto de 
			       Astrof{\'{\i}}sica de Andaluc{\'{\i}}a (CSIC).}}
\author[M. N\'u\~nez-D\'\i az et al.]
       {M. N\'u\~nez-D\'\i az\thanks{E-mail: manund@iac.es}$^{1,2}$, 
       A. Mesa-Delgado$^{1,2,3}$,  C. Esteban$^{1,2}$, L. L\'opez-Mart{\'{\i}}n$^{1,2}$,\newauthor
       J. Garc{\'{\i}}a-Rojas$^{1,2}$ and V. Luridiana$^{1,2}$ \\
        $^1$Instituto de Astrof\'\i sica de Canarias (IAC), E-38200 La Laguna, Tenerife, Spain\\
        $^2$Dept. Astrof{\'{\i}}sica, Universidad de La Laguna (ULL), E-38206 La Laguna, Tenerife, Spain\\
        $^3$ Departamento de Astronom{\'{\i}}a y Astrof{\'{\i}}sica, Facultad de F{\'{\i}}sica, Pontificia Universidad Cat{\'o}lica de Chile,\\
        $~$ Av. Vicu{\~n}a Mackenna 4860, 782-0436 Macul, Santiago, Chile\\}

\begin{document}

\date{Accepted X XX XX  Received XXXX XX XX; in original form XXXX XX XX}
\pagerange{\pageref{firstpage}--\pageref{lastpage}} \pubyear{2011}

\maketitle
\label{firstpage}

\begin{abstract}
We present results from integral field optical spectroscopy with the Potsdam
Multi-Aperture Spectrograph of the Herbig-Haro (HH) object HH 204, with a spatial
sampling of $1\times1$ arcsec$^2$.  We have obtained maps of different
emission lines, physical conditions and ionic abundances from collisionally excited lines. 
The ionization structure of the object indicates that the head of the bow shock is 
optically thick and has developed a trapped ionization front. The density at the head is at least five 
times larger than in the background ionized gas. We discover a narrow arc of high \te(\nii) values 
delineating the southeast edge of the head. The temperature in this zone is about 1,000 K 
higher than in the rest of the field and should correspond to a shock-heated zone at the 
leading working surface of the gas flow. This is the first time this kind of feature is observed in a 
photoionized HH object. 
We find that the \Op\ and O abundance maps show anomalous values at separate areas of the bow shock
probably due to: a) overestimation of the collisional de-excitation effects of the \oii\ lines in the compressed gas at the head of the bow shock, and b) the use of a too high \te(\nii) at the area of the leading working surface of the flow. 

\end{abstract}


\section{Introduction} \label{intro}
Herbig-Haro (HH) objects were originally discovered as knots of optical
emission, found in regions of low-mass star formation \citep{Herbig1951,Haro1952}. 
However, they are now observed in all spectral domains 
and are also associated with high-mass star-forming regions \citep[][]{Marti1993}.
They are often found in groups, arranged in linear or quasi-linear structures,
frequently with aligned proper motion vectors \citep{Herbig1981,Reipurth1992,
Eisloffel1994} and
are a manifestation of the collimated supersonic ejection of material
from young stars or their close circumstellar environment. HH objects
show an emission line spectrum characteristic of shockwaves, with 
velocities typically in the range $50-200$ \kms~ \citep[][]{Schwartz1980}
and, in many cases, individual knots or groups of knots have shapes that
are very suggestive of bow shocks.

There are many HH objects identified in the
Orion Nebula -- the nearest high-mass star-forming region-- which is the most observed and studied Galactic H II region. 
The most prominent high-velocity
feature in the nebula is the Becklin-Neugebauer/Kleinmann-Low (BN/KL)
complex, which contains several HH objects. In addition, there are other
important high-velocity flows that do not belong to the BN/KL complex,
as is the case of HH 202, 203 and 204. The origin of these flows has been
associated with infrared sources embedded within the  Orion-South
\citep[see][and references therein]{O'Dell2008a}. The hottest star
in the Trapezium cluster, \oric\ \citep[O7 V,][]{Simon-Diaz2006}, dominates 
the photoionization of 
the surrounding gas and produces a thin blister of photoionized material.
In general, the mixture of photoionized and shocked gas makes the interpretation 
of the spectra of HH objects within H II regions more difficult; for example, emission lines 
formed as a shock moves through ionized gas closely resemble those from
higher velocity shock in neutral gas \citep[][]{Cox1985}.
Understanding line ratios in H II regions also requires distinguishing
shocked from photoionized gas before we determine canonical 
electron temperatures and densities and ionic abundances of such regions.
\begin{figure*}
   \centering
   \includegraphics[scale=0.7]{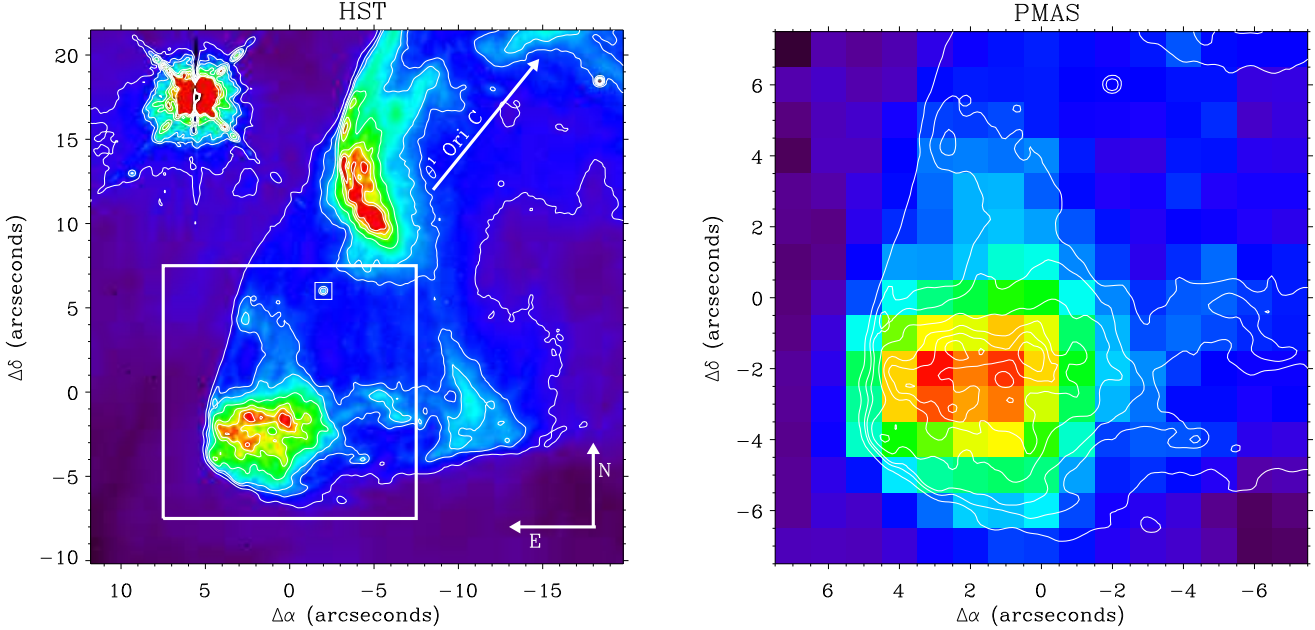}
   \caption{Left-hand side: {\em HST} image taken with the Wide-Field Planetary Camera 2 (WFPC2)
   in the narrow filter  F656N (\ha+\nii) \citep[][]{O'Dell1996}. The large white square corresponds 
   to the usable PMAS FOV covering the Herbig Haro object \hhd. Right-hand side: PMAS \ha\ emission map with
   {\em HST} contours  overplotted.}
   \label{position}
\end{figure*}	

\hhd\ --as well as the neighbouring \hhu-- is located close to \oria\
\citep[O9 V,][]{Simon-Diaz2006} and has an optical angular size of the order of tens of arcseconds.
It was discovered by \cite{Munch1962} and identified as HH object by \cite{Canto1980}.
It has been the subject of numerous subsequent
studies of their ionization structure \citep[][]{O'Dell1997b}, electron
density \citep[][]{Walsh1982}, radial velocity \citep[][]{O'Dell1993}
and tangential motions \citep{Cudworth1977,Hu1996,Doi2002}.
\cite{Rosado2002}, in a kinematic study based on Fabry-P{\'e}rot data, 
suggested that it seems to be part of a large structure or lobe.
Also, these authors showed that the \oiii~emission does
not come from the head of the bow shock, but instead occurs primarily in a
diffuse region along the southwestern portion of the object. \cite{O'Dell1997a}
attribute the distribution of \oiii~to differences in the way in which \oric\
illuminates this region of the nebula. The ultraviolet light from this star
penetrates into the back side of the shock which moves toward the observer 
\citep[][]{O'Dell1997b}

HH 204 is a highly filamentary bright knot at the end of a rather symmetric 
cone that resembles a bow shock. \cite{O'Dell1997a} indicate that the excitation in this object appears to be
controlled by a mixture of local shocks and external ionization from 
\oric. It presents a pronounced asymmetry in brightness
between the bow shock side near \oria\ and the side
located away from this star, which is better appreciated in \nii\ than in \ha.
\cite{Henney1996} has proposed that a transverse density gradient in the ambient
medium where a bow shock propagates could lead to an asymmetry in brightness
of the bow shock. This idea is supported by \cite{Rosado2001}, who found
a slight velocity gradient running perpendicular to the axis of HH 204 in the 
sense that the fainter (and less dense) regions show larger velocities.
 Originally, \cite{O'Dell1997b} interpreted that the shock of 
\hhd\ was formed when the jet hits neutral material in the foreground Veil
\citep[][]{O'Dell2001}, but more accurate knowledge of the position
of Orion's Veil \citep[][]{Abel2004} and an improved determination of the 
trajectory of the shock allowed \cite{Doi2004} to establish that \hhd\ arises
where the jet strikes denser nebular material located where the main ionization front 
of the Orion Nebula tilts up. This tilt is what gives rise to the Bright Bar feature that 
runs from northeast to southwest, passing slightly at the northwest of \oria. 

In previous papers, our group has studied the spatial distribution and properties of the nebular gas at small spatial scales in the 
Orion Nebula. Using long-slit spectroscopy at spatial scales of $1.2$ arcsec and crossing different
morphological structures, \cite{Mesa-Delgado2008} found spikes in the distribution of the electron
temperature and density which are related to the position of proplyds and HH objects. In addition, we have published two  
additional studies using integral-field spectroscopy at spatial scales of $1\times1$ arcsec${^2}$, 
focused on the analysis of certain interesting morphological structures. \cite{Mesa-Delgado2009b} studied 
the prominent HH 202 and \cite{Mesa-Delgado2011} analysed areas of the Bright
Bar and the northeast of the Orion South Cloud. They have mapped the distribution of the emission line fluxes and 
the physical and chemical properties of the nebular gas in the selected areas.

The main goal of this paper is to use integral field spectroscopy at spatial scales of 1 arcsec in order to explore 
the ionization structure, physical conditions and chemical abundances of one of the most remarkable HH object of the Orion
Nebula: \hhd.  

In Section 2 we describe the observations obtained and the reduction procedure. In Section 3 we describe the 
differential atmospheric refraction correction, the emission-line measurements and the reddening correction
of the spectra as well as the determination of the physical conditions and chemical abundances.
In Section 4 we present the spatial distribution of emission line ratios, physical conditions and chemical
abundances. In Section 5 we discuss the trapped ionization front and the high-temperature
arc. Finally, in Section 6 we summarize our main conclusions.

\section{Observations and Data Reduction} \label{obsred}
The observations were made on 2008 December 20 at the 3.5m Calar Alto telescope
(Almer\'\i a, Spain) with the Postdam Multi-Aperture Spectrometer \citep[PMAS,][]{Roth2005}
in service mode. The standard lens array integral field unit (IFU) 
was used with a sampling of 1$\arcsec$ and $16\arcsec\times16\arcsec$ field of view (FOV).
The blue range ($3500$ to $5100$ \AA) and red range ($5700$ to $7200$ \AA) were covered with the 
V600 grating using two grating rotator angles: $-72$ and $-68$, respectively, with an effective 
spectral resolution of $3.5$ \AA. The blue and red spectra have a total integration time of
$2400$ and $300s$, respectively.
Additional short exposures of $30s$ were taken in order to avoid saturation of the 
brightest emission lines.
Calibration images were obtained during the night: continuum lamp needed to extract the 256
apertures, arc lamps for the wavelength calibration and spectra of the spectrophotometric
standard stars Feige 34, Feige 110 and G~191-B2B \citep{Oke1990} for flux calibration.
The error of the absolute flux calibration of the spectra is of the order of 5\%.  The observing
night had an average seeing of about $1.5\arcsec$.

The data were reduced using the {\sc iraf} reduction package {\sc specred}. After bias 
subtraction, spectra were traced on the continuum lamp exposure obtained before and after 
each science exposure, and wavelength calibrated using a Hg--Ne arc lamp. The dome flats were 
used to determine the response of the instrument for each fiber and wavelength. Finally, for the 
standard stars we have co-added the spectra of the central fibers and compared them with the 
tabulated one-dimensional spectra.

On the left-hand side of Fig. \ref{position}, we show a high resolution Hubble Space Telescope ({\em HST}) image
taken with the Wide-Field Planetary Camera 2 (WFPC2) in the narrow-band filter F656N (\ha\ + continuum) 
retrieved from its public archive. Two images were combined using the task {\em crrej} of the 
{\em hst\_calib.wfpc} package of {\sc iraf} to reject cosmic rays. 
To locate precisely the PMAS FOV on the {\em HST} image, we have used the star marked with a small 
white box in Fig. \ref{position} (left), which is visible in the 
spaxel $(-2,6)$ in all the continuum maps of the PMAS field. Therefore, the accuracy in the location of the usable PMAS FOV (the large
white box on the {\em HST} image) is better than $1\arcsec$. With the {\em HST} image, we computed high-resolution contours and    
overplotted them on all maps. Note that the resolution of the {\em HST} image is roughly one order of magnitude better than the PMAS one 
(see right-hand side of Fig. \ref{position}).

\section{Analysis} \label{medpmas}
\subsection{DAR correction}
\begin{figure}
   \centering
   \includegraphics[scale=0.45]{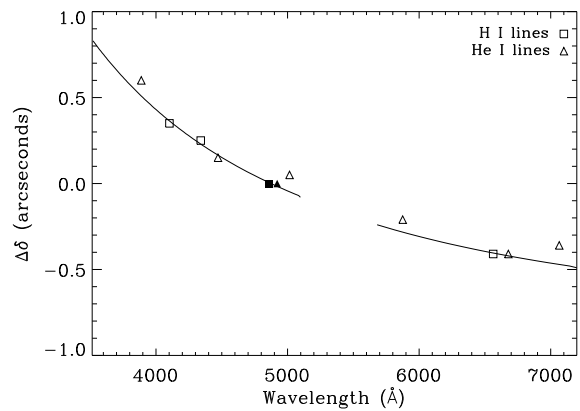} 
   \caption{Comparison between the differential atmospheric refraction (DAR) values computed from Filippenko's model (solid lines) and the
     declination shifts obtained from our data using the {\sc correl\_optimize} routine avaible in the {\sc IDL} 
     Astronomy User's Library. The squares represent the shifts of the \ion{H}{i}
     lines with respect to \hb\ (filled) and the triangles the \ion{He}{i} lines with
     respect to \ion{He}{I} 4921 \AA\ (filled). The gap in the continuous line shows 
     the part of the optical spectrum do not covered with the blue and red ranges of our observational setup. 
    The symbols in the red range have been shifted $-0.20$ arcseconds (see text).}
   \label{dar}
\end{figure}
The effect of the differential atmospheric refraction (DAR) can be noticed in the data. 
We have used the monochromatic images of \ion{H}{i} and \ion{He}{i} emission lines at different 
wavelengths --using \hb\ and \ion{He}{I} 4921 \AA\ as references-- to correct for this effect. 
With the {\sc correl\_optimize} routine avalaible in the {\sc IDL} 
Astronomy User's Library we computed the optimal pixel offset of the monochromatic image of each selected line relative 
to its reference by maximizing the correlation function. In Figure \ref{dar} we represent the 
displacements in declination due to the DAR compared with the results of Filippenko's model 
\citep[][]{Filippenko1982} and with respect to H$\beta$ and \ion{He}{I} 4921 \AA. As the optical spectra analyzed are separated in two 
spectral ranges, there is an additional fixed instrumental shift between the lines of the red and blue ranges. This shift 
has been double checked using the pair of images \ha$-$\hb\ and \ion{He}{i} 
$4921-6678$ \AA, yielding displacements of $(0.10,-0.20)$ and $(0.10,-0.25)$ pixels, respectively. We chose the first 
pair of values because it is computed with much brighter lines. Therefore, the displacements of lines in the red range 
of the spectrum have been shifted $(0.10,-0.20)$ arcsecs. Figure \ref{dar} only presents
the behaviour of DAR in the declination axis because it has larger values than in right ascension 
due to the paralactic angle is better aligned with the declination axis.
Attending to the good agreement of the model with the data values, 
we corrected from DAR the whole set of data
cube applying the {\sc bilinear} routine of {\sc idl} with a global and independent offset to each
axis, in order to maximize the coincident FOV ($15\times15$ arcsec$^2$).
\subsection{Line intensity measurements}
The emission lines considered in our analysis are the following: hydrogen Balmer lines, from \ha\ to
H12, which are used to verify the DAR correction and to compute the reddening coefficient; collisionally 
excited lines (CELs) of various species, which are used to compute physical conditions and chemical abundances.

Line fluxes were measured applying a single or a multiple Gaussian profile fit procedure between
two given limits and over the local continuum. All these measurements were made with the {\sc 
splot} routine of {\sc iraf} and using our own scripts to automatize the process. The associated errors 
were determined following \cite{Mesa-Delgado2008}. In order to avoid spurious weak line measurements, we imposed three 
criteria to discriminate between real features and noise: 1) Line intensity peak over 2.5 times 
the sigma of the continuum; 2) FWHM(\hi)/1.5~$<$~FWHM(\ld)~$<$~1.5$\times$FWHM(\hi); and 3) F(\ld)~$>$~0.0001~$\times$~F(\hb).\\
All line fluxes of a given spectrum have been normalized to \hb\ and \ha\ for the blue 
and red range, respectively. To produce a final homogeneous set of line ratios, all of them were 
re-scaled to \hb. The re-scaling factor used in the red spectra was the theoretical 
\ha/\hb\ ratio for the physical conditions of \te~=~10000 K and \nel~=~1000 \cmc.
\begin{figure}
   \centering
   \includegraphics[scale=0.75]{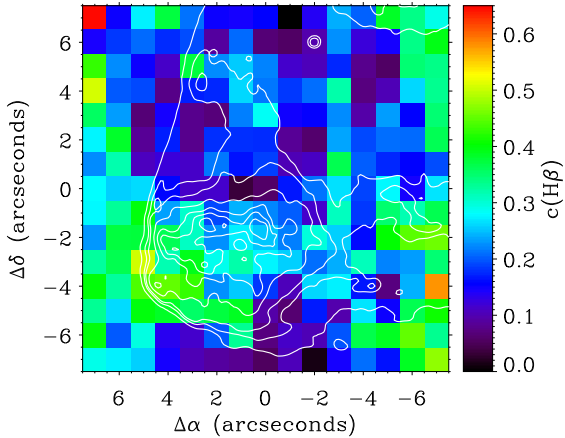} 
   \caption{Spatial distribution map of the extinction coefficient,
   c(\hb).  The contours plotted are those from the {\em HST} F656N images.
   The size of the field corresponds to the maximal coincident FOV after correcting the spectra 
   for differential atmospheric refraction (see text).}
   \label{chb}
\end{figure}
\subsection{Reddening coefficient}
The reddening coefficient, c(\hb), has been obtained by fitting the observed 
\hg/\hb\ and \hd/\hb\ ratios to the theoretical 
ones predicted by \cite{Storey1995} for \nel~=~1000~\cmc\ and \te~=~10000~K. We have used 
the reddening function, $f(\lambda)$, normalized to \hb\ determined by \cite{Blagrave2007} 
for the Orion Nebula. The use of this extinction law instead of the classical one of 
\cite{Costero1970} produces c(\hb) values about $0.1$~dex higher and dereddened fluxes 
with respect to \hb\ about 3\% lower for lines in the range 5000 to 7500 \AA\ and 4\% higher for 
wavelengths below 5000 \AA. The final adopted c(\hb) value for each spaxel is an average 
of the individual values derived from each Balmer line ratio weighted by their corresponding
uncertainties. The typical error is about $0.1$~dex for each spaxel. 
The resulting extinction map is shown in Figure~\ref{chb}, where the extinction
coefficient varies approximately from $0.1$ to $0.6$ dex.
The mean value in the FOV is $0.23$~dex. 
We have compared our c(\hb) value with the extinction map obtained by \cite{O'Dell2000} 
from the \ha/\hb\ line ratio using calibrated {\em HST} images, where \hhd\ is located
between the contour lines of $0.2$ and $0.4$ dex on that map.
Therefore, we find that our mean value of the reddening coefficient is consistent with the 
previous determinations taking into account the errors and the different extinction laws used. 
 However, we also find some regions with higher values of extinction than those reported by \cite{O'Dell2000}.
It must to be said that, due to its closeness to \oria, the \hd\ and \hg\ lines of the spectra  
are somewhat affected by underlying absorption due to the stellar continuum reflected by dust. 
This may affect the measurement of the intensity of Balmer lines, especially the bluest ones, 
producing a slightly higher c(\hb), and this may be the reason of such discrepancy.
However, the results of our paper are not particularly sensitive to this issue.
\subsection{Physical conditions}
Nebular electron densities, \nel, and temperatures, \te, have been
derived from the usual CEL diagnostic ratios --\sii~$6717/6731$ for \nel,
and \nii\ $(6548$+$6583)/5755$, \sii\ $(6716$+$6731)/(4068$+$4076)$ and \oiii\ $(4959$+$5007)/4363$ 
for \te-- and using the {\sc temden} task of the {\sc nebular} package of 
{\sc iraf}  with updated atomic data \citep[see][]{Garcia-Rojas2009}. 
Following the same procedure as \cite{Mesa-Delgado2008}, we have 
assumed an initial \te\ = 10000 K to derive a first approximation of \nel(\sii), then we calculate 
\te(\nii), and iterate until convergence. We have not corrected the 
observed intensity of \nii\ 5755 \AA\ for contribution by recombination
when determining \te(\nii) because this is expected to 
be rather small in the Orion Nebula \citep[e.g.][]{Esteban2004}. 
We did not use \te(\sii) for the iteration procedure because the auroral lines 
of \sii\ are rather weak outside \hhd\ and the diagnostic line ratio involves 
lines located in the blue and red spectral ranges. Nevertheless, using both \sii\ 
diagnostic ratios together the density values are in good agreement within the errors 
with those determined with \te(\nii). On the other hand, because of the
relatively low spectral resolution of the observations, \oiii\ 4363 \AA\ is somewhat
blended with the \ion{Hg}{i}+\feii\ 4358 \AA\ lines. The latter one becomes very strong inside 
the HH object. Only in this case, we performed manually the Gaussian fit of those three lines 
in order to obtain a good determination of \te(\oiii). The automatic procedure gave some inconsistent 
values at the northeast corner of the FOV because of the faintness of the 
\oiii\ 4363 \AA\ line in that area. 
The typical uncertainties in the density maps range from $200$ to
$2000$ \cmc. The errors for electron
temperatures range from $150$ to $350$ K for \te(\nii), from $100$ to $400$ K for
\te(\oiii) and from $100$ to $500$ K for \te(\sii). 
\subsection{Chemical abundances}
The {\sc iraf} package {\sc nebular} has been used to derive ionic abundances of \Sp, \Np, \Op\ 
and \Opp\ from the intensity of CELs. We have assumed no temperature fluctuations in the ionized 
gas ($t^2=0$) and a two-zone scheme: \te(\nii) is used to derive the \Sp, \Np\ and \Op\ --the low 
ionization potential ions--  and \te(\oiii) for the \Opp\ abundances. The uncertainties in ionic 
abundances have been calculated as a quadratic sum of the independent contributions of errors in 
\nel\ and \te.  For oxygen abundances, the uncertainties are $0.04-0.08$ dex for 
\Op, $0.02-0.05$ dex for \Opp\ and $0.02-0.06$ dex for O.

\section{Results: Spatial distribution maps}
In this section we analyse the spatial distributions of several emission line
ratios, electron densities and temperatures as well as several
representative ionic and total abundances.

\subsection{Ionization structure}\label{res_ion_struc}
\begin{figure*}
   \centering
   \includegraphics[scale=1.0]{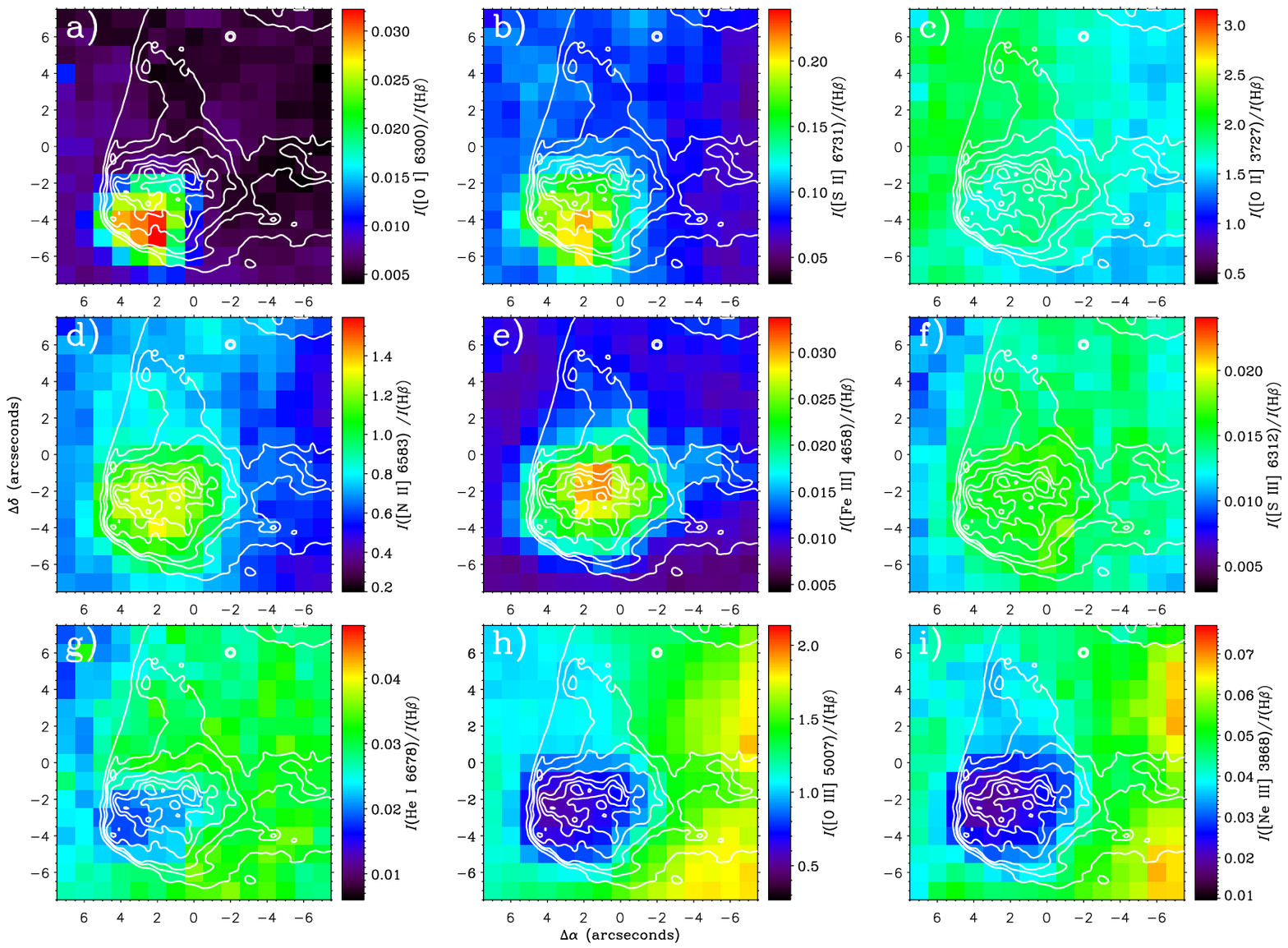}
   \caption{Maps of line intensity ratios relative to \hb\ of selected emission lines: a) \oi\ 6300 \AA, b)\sii\ 6731 \AA,
  c) \oii\ 3727 \AA, d) \nii\ 6583 \AA, e) \feiii\ 4658 \AA, f) \siii\ 6312 \AA, g) \hei\ 6678 \AA, h) \oiii\ 5007 \AA\
  and i) \neiii\ 3868 \AA.  The contours plotted are those from the {\em HST} F656N images.
  The color bars have been re-scaled to have the same dynamical range
  between the maximum and minimum values in all maps.}
   \label{fluxes204}
\end{figure*}
In Fig. \ref{fluxes204}, we show the line intensity ratio maps 
of \oi\ 6300 \AA, \sii\ 6731 \AA, \oii\ 3727 \AA,
\nii\ 6583 \AA, \feiii\ 4881 \AA, \siii\ 6312 \AA, \hei\ 6678 \AA, \oiii\ 5007~\AA\ and \neiii\ 3868 \AA\ 
relative to \hb.
The minimum and maximum values of the color bars of the different ratios 
have been rescaled to show the same dynamical range in all the maps. 
The dynamical range of reference adopted is that of the \oi/\hb\ ratio, which shows the 
largest variation. 
The selected lines cover all possible ionization conditions of the gas. 
The maps are sorted from lower (top-left) 
to higher (bottom-right) ionization potential. These line ratios are directly related to the local
ionization degree of the gas and do not depend on density in a direct manner.

\oi\ 6300 \AA\  is a strong line in shocks, but is rather weak in photoionized regions 
\citep[e.g.][]{Masciadri2001}. However, this line can also be a tracer of the presence of trapped ionization
fronts \citep[as in the case of HH~202, in the Orion Nebula:][]{Mesa-Delgado2009a,Mesa-Delgado2009b}. 
In Fig. \ref{fluxes204}a, we can see that \oi/\hb\ is much higher in the southeast quadrant 
of the bow shock and very low along the bow shock wings. This indicates that the wings of the flow 
are completely photoionized. In order to disentangle the excitation mechanism of the gas at the bow shock, 
we have represented its \oi/\ha\ and \oi/\oiii\ ratios on a diagnostic diagram adapted from 
\cite{Riera1990} that separates shock excitation from photoionization, finding that it is esentially photoionized, 
and that no substantial contribution of shock excitation is needed to reproduce its spectrum. In their models of 
optically thick HH objects inside photoionized regions, \cite{Masciadri2001} find that trapped ionization 
fronts restricted at the head of the bow shock show an \oi/\hb\ ratio concentrated to the side 
directed away from the ionizing photon source, exactly what we see in Fig. \ref{fluxes204}a. The \sii 
~line emission is also a tracer of ionization fronts. In Fig. \ref{fluxes204}b, we can see 
that \sii/\hb\ follows rather closely the distribution of \oi/\hb. 

As we can see in Fig. \ref{fluxes204}c and \ref{fluxes204}h, the distribution of \oii/\hb\ 
shows an almost inverse behaviour with respect to \oiii/\hb\, presenting its lowest values at the 
western part of the FOV, exactly where the \oiii/\hb\ ratio is higher. However, \oiii/\hb\ reaches its minimum 
at the head of the bow shock while \oii/\hb\ does not present  peak values at those spaxels. This indicates 
that the spatial distribution of \oii\ 3727 \AA\ and \hb\ is rather similar at the bow shock. 

The \nii/\hb\ ratio (Fig. \ref{fluxes204}d) reaches its maximum at 
the bow shock, and delineates the wings of the flow structure. It can also be seen that the 
area closer to the star \oria\ is brighter than the rest. This result is in agreement with 
\cite{Rosado2001}, who reported that the asymmetry in the brightness distribution of \hhd\  
is better appreciated in \nii\ than in \ha. 

HH objects are also characterized by their strong emission in \feiii\ lines. In Fig. 
\ref{fluxes204}e, we can see the spatial distribution of the \feiii/\hb\ ratio where
the maximum values seem to be concentrated just behind the areas of the maxima of the \oi/\hb\ and \sii/\hb\ ratios.
The high \feiii\ emission may be related to destruction of dust grains due to the passage of shock waves 
as in the case of HH~202 \citep{Mesa-Delgado2009a,Mesa-Delgado2009b}. 

The spatial distributions of \siii/\hb\ and \hei/\hb\ are rather similar and smooth (Fig.
\ref{fluxes204}f, g). Higher values tend to be concentrated toward the west of the FOV. 
This concentration is more evident in the higher ionized species \oiii/\hb\ and \neiii/\hb\  
(Fig. \ref{fluxes204}h, i) due to the contrast between the minimum and maximum line 
ratios is more pronounced. It is important to comment that the high excitation lines of \oiii\ and \neiii\ do not form 
at the head of the bow shock like they usually do in other HH objects produced by high-velocity flows, 
but instead occur primarily in the ambient diffuse background across the western portion of the FOV. This distribution is attributed to 
differences in the way in which the star \oric\ illuminates this zone of the nebula \citep{O'Dell1997b} and the relatively 
large distance between the star and \hhd, which makes the impinging ionizing flux too low to 
maintain the presence of a substantial fraction of \Opp\ ions at the bow shock. 

\subsection{Physical conditions}\label{res_phys_cond}

\begin{figure*}
   \centering
   \includegraphics[scale=0.95]{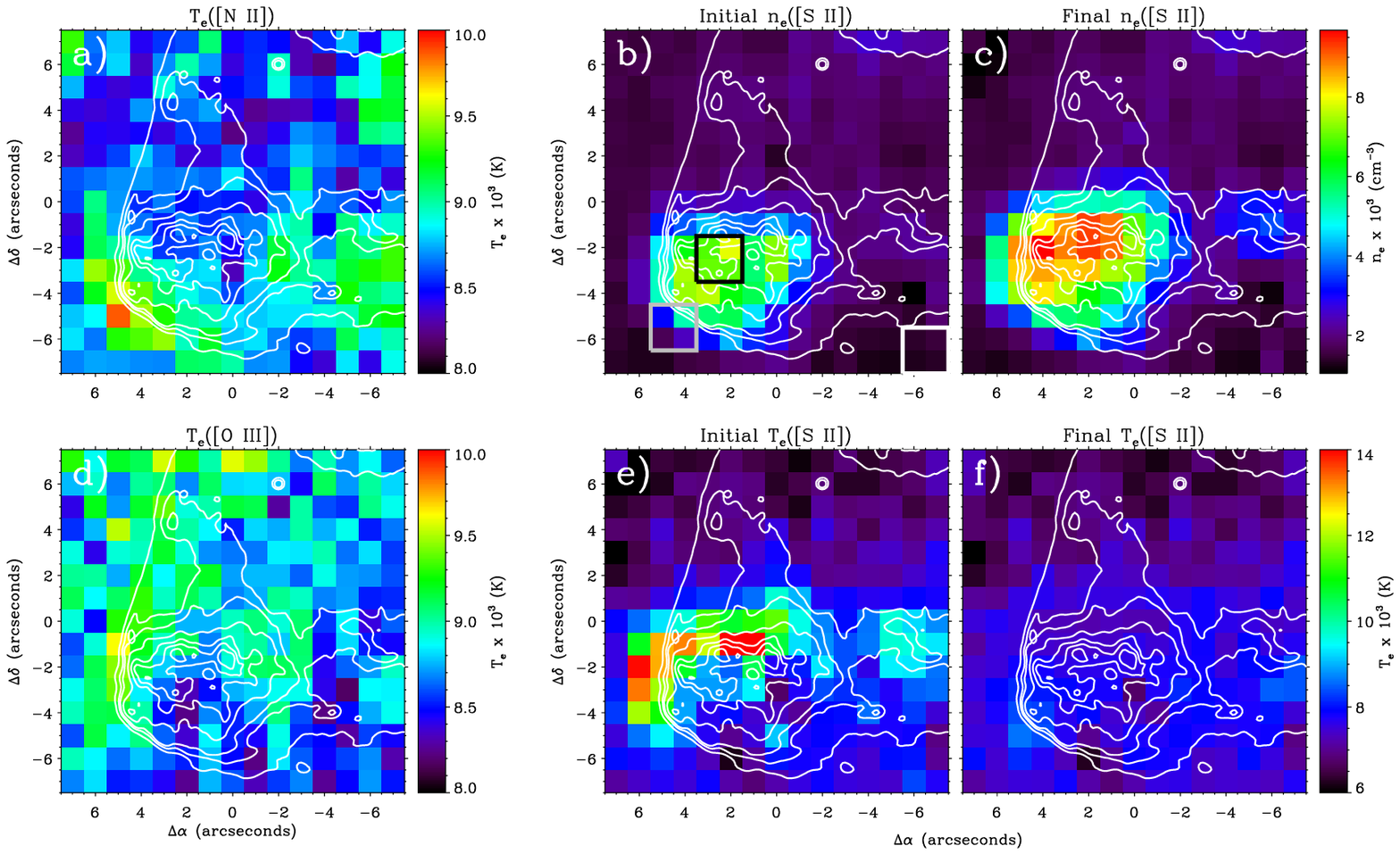}
   \caption{Physical conditions of \hhd\:
   a) \te(\nii), b) and c) two different estimates of \nel(\sii), d) \te(\oiii), e) and f) two different estimates of \te(\sii). 
   See text for details.  The contours plotted are those from the {\em HST} F656N images. 
   In the density map b) we mark three different zones: background gas 
   (white box), head of the bow shock (black box) and high-temperatutre arc (grey box), whose spaxels have been co-added in order to analyse their spectra and 
  derive the physical conditions at those zones.}
   \label{condfis204}
\end{figure*}

In Fig. \ref{condfis204} we present the physical conditions determined for the observed field,
where the head of the bow shock is associated with local peaks of electron density. 

Fig. \ref{condfis204}a shows the \te(\nii) map. A remarkable feature is the narrow high-temperature zone just at the southeast border 
of the head of the bow shock. This structure seems to be real, since the temperature determination has only an error of about 2\%, much lower than 
the temperature increase. Furhtermore, it is a rather narrow feature that can only be distinguished when the DAR correction is applied.  
 On the other hand, this high-temperature zone cannot be produce by the slight extinction excess we find in a nearby --but not coincident-- area
of the head. We would need c(\hb) differences by about unity to account for a temperature increase of 1000~K.
We also do the exercise of correcting the whole set of data with two constant values of c(\hb), 0.2 and 0.4 dec, and the high-temperature arc is still
clearly present in both cases. Then the derived temperature enhancement is not an artifact of the high local extinction.
The nature of this structure will be discussed later. Fig. \ref{condfis204}b shows the \nel(\sii) map derived in combination with \te(\nii) 
until convergence. Assuming this initial \nel(\sii) map we also derived the \te(\sii) map (see Fig. \ref{condfis204}e), 
which shows an arc that delineates a zone of strong increase of temperature at the northeast of the head. 
We found that this structure also appears if we calculate the map of the \ioni{S}{+} abundances derived from 
the auroral \sii\ lines when using \te(\nii) (Fig. \ref{condfis204}a) and \nel(\sii) (Fig. \ref{condfis204}b).  
Obviously, that result has no physical meaning. The simpler explanation for this fact is that we are not using appropriate 
values of the  electron density for deriving \ioni{S}{+} abundances and this affects especially the calculations based on nebular \sii\ lines, which have lower critical 
densities than the auroral ones. 
In order to correct for this discrepancy in the \ioni{S}{+} abundance, we determined the \nel(\sii) map that --assuming 
the \te(\nii) distribution of Fig. \ref{condfis204}a-- makes the differences between
the abundance obtained from the nebular and auroral lines lower than the estimated maximum error of this 
abundance, $\pm$0.08 dex. The final electron density map is shown in Fig. \ref{condfis204}c, where only half of the spaxels
show values different from the previous map. It must be said that using  an iterative procedure in which 
both \te(\nii) and \nel(\sii) are changed we do not find a substantially different final \te(\nii) map because the 
\nii\ lines involved in the temperature indicator have critical densities much higher than the \nel(\sii) values we are obtaining 
for \hhd. 
In Fig. \ref{condfis204}c we can see that this new density map fits much better
the \ha\ contours not only inside the head of the bow shock, but also at the location 
of other structures visible in the {\em HST} images such as the wing at the west of the head. 
We recomputed \te(\sii) assuming the \nel(\sii) map of Fig. \ref{condfis204}c and obtained 
the distribution showed in 
Fig. \ref{condfis204}f, finding a much smoother map where the arc-shaped area of very high temperatures has 
disappeared. The final \nel(\sii) map of Fig. \ref{condfis204}c is the one we have used to derive the chemical abundances. 

In Fig. \ref{condfis204}c we can see that the highest densities in the FOV  
are between $4000$ and $9000$~\cmc. Similar values were obtained by \cite{Mesa-Delgado2008}
from a slit crossing \hhd\ and roughly parallel to the direction toward \oric\ (see Fig. \ref{rend}). 
The density map shows the highest values at the roundish clumpy structure of the head of the 
bow shock while the extended wings show densities only slightly higher than those of the surrounding 
background gas. 
The lowest density values (\nel$<2000~$\cmc), related to the background gas, 
are located at the southwest edge of both FOV. 
We obtained an additional density determination making use of the \feiii\ lines. 
We detected at least 5 \feiii\ lines of the 3F multiplet (\feiii\ 4607,4658, 4702, 4734 and 4755 \AA) 
and one line of the 2F multiplet (\feiii\ 4881 \AA) at those spaxels at the head of the 
bow shock showing the maximum values of  \nel(\sii).  The range of validity of this density indicator, 
\nel(\feiii), has an upper limit well above $10^6$ \cmc. 
To carry out these calculation we have used a 34-level atom using the 
collision strengths from \cite{Zhang1996} and the transition probabilities of 
\cite{Quinet1996} and \cite{Johansson2000}. The calculations of \nel(\feiii)
give densities varying from 8000 to 15000 \cmc, but the uncertainties are rather large 
due to the faintness of the \feiii\ lines. Therefore,
the \nel(\sii) values we obtain in Fig. \ref{condfis204}c should be considered a fair approximation to  
the true electron density of the ionized gas at \hhd. It is interesting to comment that \cite{Solf1988} found a rather 
similar lack of correlation between auroral and nebular \sii\ lines in a one-dimensional spectrum across the head of 
the prototypical HH~1. They also found rather high \nel(\sii) values --of the order of 3000~\cmc-- that peak at the head, 
precisely where the intensity ratio between auroral and nebular \sii\ lines is higher. This behavior is also similar 
to that observed in our case and can also be explained by the effect of collisional deexcitation of the nebular lines. 

The spatial distribution of \te(\oiii) (Fig. \ref{condfis204}d)
is rather smooth, with higher values located at the northeast edge of the FOV. 
This area shows the faintest \oiii~4363~\AA\ emission and therefore have 
the largest errors. In fact, the anomalously high \te\ of some spaxels might indicate that the
\oiii~4363~\AA\ line has not been properly deblended from Hg~I+\feiii~4358~\AA\ lines.
There is an apparent trend of slightly higher \te(\oiii) around the northern half 
of the head of the bow shock. 

\begin{table*}
 \centering
  \caption{Physical conditions of the three selected zones (see Fig. \ref{condfis204}b).}
  \begin{tabular}{@{}r|c|c|c|c|c|c@{}}
  \hline
  Zone               &  c(\hb)      & \nel(\sii)   & \nel(\feiii)    & \te(\nii)     & \te(\sii)    &  \te(\oiii)  \\ 
                     &  (dec)       & (\cmc)       & (\cmc)          & (K)           & (K)          & (K)          \\\hline \hline
  background         & $ 0.4\pm0.1$ & $1400\pm300$ & $2800\pm1900$   &  $8800\pm300$ & $7700\pm300$ & $8500\pm200$ \\
  head               & $ 0.3\pm0.1$ &$9000\pm1800$ &  $15300\pm5600$ &  $8700\pm150$ & $7800\pm250$ & $8700\pm350$ \\
high-temperature arc & $ 0.3\pm0.1$ & $3800\pm550$ & $5000\pm500$    & $9600\pm200$  & $8500\pm300$ & $8800\pm250$ \\

  \hline
\end{tabular}
\label{zones}
\end{table*}

In Fig. \ref{condfis204}b we mark three selected areas covering different zones, whose spectra have been extracted in 
order to analyse their physical conditions. These areas are: background gas (white box), head of the bow shock (black box) and  
high-temperature arc (grey box). This last zone corresponds to the area showing the peak of \te(\nii) located 
at the southeast of the head (see Fig. \ref{condfis204}a). 
In Table \ref{zones}, we summarize the density and temperature values we obtain for each zone. The
\nel(\sii) of the three zones are very different, 1400~\cmc\ for the background, 9000~\cmc\
for the head and 3800~\cmc\ for the high-temperature arc. It is remarkable that the electron densities determined with
\feiii\ lines are rather consistent with the quoted values of \nel(\sii) within the uncertainties and show the same qualitative behaviour.  
On the other hand, the electron temperature of low ionization species shows similar values in the background and the head of the bow shock 
and higher ones at the high-temperature arc.  In the case of \te(\nii) the increase is of about 1,000~K and 
in the case of \te(\sii) it is somewhat lower, about 700~K.
Finally, \te(\oiii) is practically the same in the three areas within the errors. 
These results indicate that there might be a process of extra heating in the high-temperature arc area (see Section \ref{high_t_arc}). This process only affects 
to low-ionization ionic species, because --attending to the results of Section \ref{res_ion_struc} and Section \ref{chem_abund}-- high-ionization species --such as \Opp-- are absent at the head of \hhd.

\subsection{Chemical abundances}\label{chem_abund}

\begin{figure*}
   \centering
   \includegraphics[scale=1.]{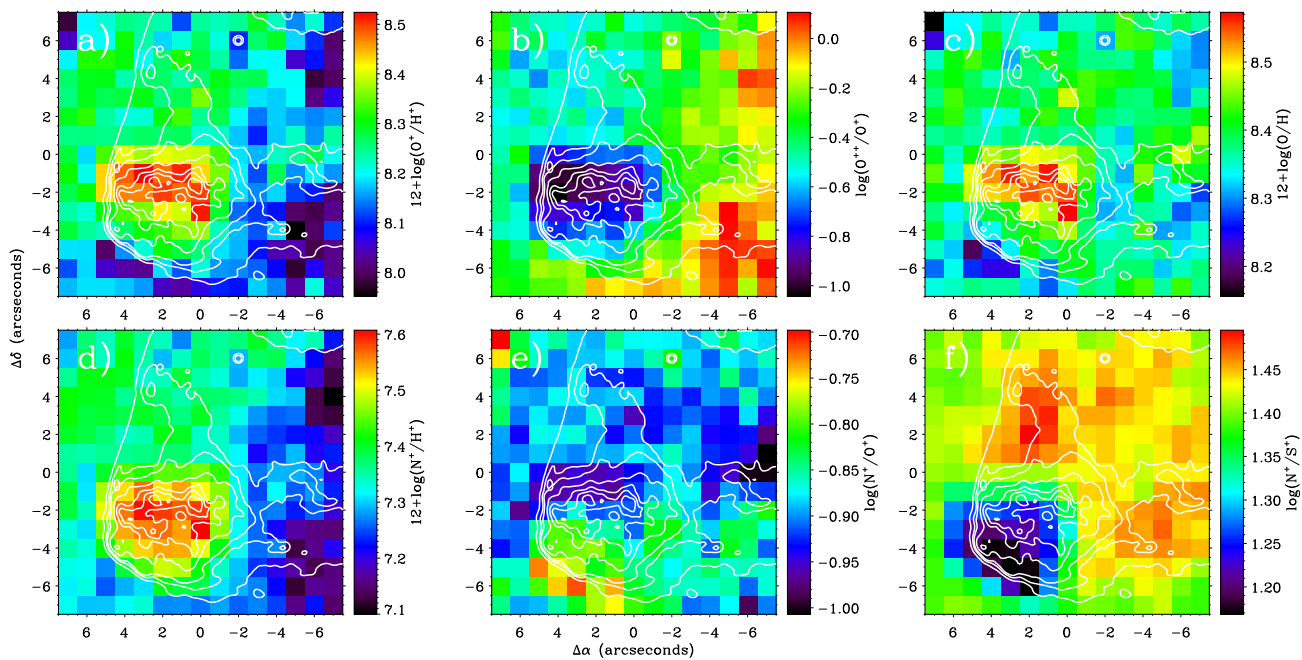} 
   \caption{Ionic abundance and abundance ratio maps determined from the intensity of collisonally excited lines. 
    From top-left to bottom-right: 12 $+$ log(\Op/\Hp), log(\Opp/\Op), 12 $+$ log(O/H), 12 $+$ log(\Np/\Hp), log(\Np/\Op) and log(\Np/\Sp).
    The contours plotted are those from the {\em HST} F656N images.}
   \label{abund204}
\end{figure*}

In the first row of Fig. \ref{abund204} we present the spatial distributions of
the  \Op/\Hp, \Opp/\Op\ and O/H ratios derived from CELs.
The \Op\ abundance is strongly concentrated at the north of the head (see Fig. \ref{abund204}a). 
As we can see in Fig. \ref{abund204}b, the \Opp/\Op\ ratio shows an almost 
inverse behaviour with respect to \Op\ abundance, indicating the strong ionization stratification across the FOVs.
Unexpectedly, we do not find a constant value in the spatial
distribution of the total O abundance, as we can see in Fig. \ref{abund204}c. 

The mean value of the O abundance amounts to 12+log(O/H) = $8.40\pm0.04$ dex, which
is slightly lower than the typical values of about 8.50 dex obtained by previous
works in other parts of the Orion Nebula 
\citep[][]{Esteban1998,Esteban2004,Blagrave2006,Mesa-Delgado2009a,Simon-Diaz2011}. 
The maximum values --with abundances between 0.1 and 0.2 dex higher-- are concentrated at 
the north half of the head where the errors are minimal and the minimum values --about 0.15 dex lower-- are just 
at the high-temperature arc. It is remarkable that the \Op\ (Fig. \ref{abund204}a) and 
O (Fig. \ref{abund204}c) abundance maps are almost identical, indicating that the variations of the total O abundance are produced by 
the structure of the \Op\ abundance. A similar result has also been obtained in a previous paper of our group devoted to the 
analysis of PMAS data of ionization fronts in the Orion Nebula \citep{Mesa-Delgado2011}. However, in that paper and in contrast to our results for \hhd, we find that the \Op\ and O maps were also similar to that of \nel(\sii). 
As we can see in Fig. \ref{abund204}c, the \Op\ and O maps reach a minimum just at the southeast edge of the head of the 
bow shock --just at the position of the \te(\nii) peak of Fig. \ref{condfis204}a-- and this feature does not have any counterpart 
in neither the initial nor the final \nel(\sii) maps (Figs. \ref{condfis204}b and c, respectively). 

Considering all the aforementioned results, we propose that two different physical effects are affecting the determination of the \Op\ abundance --and consequently, the total O abundance--  across the field of \hhd. First, following the arguments outlined in \cite{Mesa-Delgado2011}, the most likely explanation for the higher \Op/\Hp\ values obtained 
at the head is that the values we obtain for \nel(\sii) are higher than the true ones of the \Op\ zone. Therefore, we are overestimating 
the collisional de-excitation of the \oii\ $3727$ \AA\ lines. Taking into account the 
\Op/\Hp\ excess at the head of \hhd\ and the results of \cite{Mesa-Delgado2011} we estimate that the true densities at the \Op\ zone should be 
about 1000 \cmc\ lower than those indicated by \nel(\sii). If we correct the density by that factor --which is of the order of 
the quoted errors-- we obtain an O abundance similar to that of the background gas around \hhd. Second, 
the effect producing the slightly lower \Op\ abundances 
at the high-temperature arc seems to be connected with the localized increase of \te(\nii) seen in Fig. \ref{condfis204}a. 
Assuming that the unexpected abundance pattern is the consequence of using an electron temperature which is 
higher than that expected considering thermal equilibrium of a static photoionized nebula, we have estimated 
by which amount we have to correct the \te(\nii) values of each spaxel of the high-temperature arc  
to reproduce the mean nominal value of the total O abundance, 8.40 dex. The result is that, by decreasing \te(\nii) by 1,000 K, we obtain 
the expected O abundance and the arc of high \te(\nii) disappears. This result support the idea that
there might be an extra heating due to shocks.

Finally, we present the \Np/\Hp, \Np/\Op\ and \Np/\Sp\ ratios in the bottom row of Fig. \ref{abund204}. The \Np\ 
abundance shows a rather similar spatial distribution to that of \Op\, as it is expected from their similar ionization
potentials. However, the ratio between both ionic abundances shows slight spatial variations concentrated 
at the south edge of the head. These variations can be due to the different temperature 
and density dependence of the lines used to derive both ionic abundances. Finally, the \Np/\Sp\ ratio delineates the presence of a trapped ionization front 
at the head of the bow shock and the strong ionization gradient across the FOV.

\section{Discussion}

\begin{figure*}
   \centering
   \includegraphics[scale=0.85]{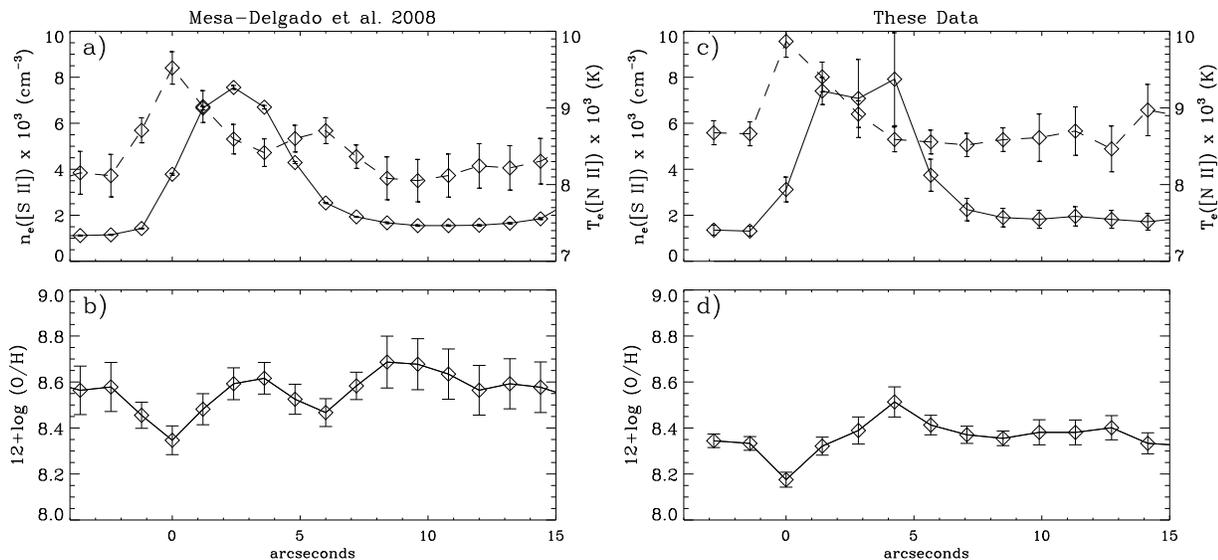}
   \caption{Spatial profiles of several parameters along slit position 3 of
   \citet{Mesa-Delgado2008} crossing \hhd\ on the left panels and a pseudo-slit along the main diagonal of our
   PMAS maps on the right.  Top panels: profiles of \nel(\sii) ({\em solid line}) and \te(\nii) 
   ({\em dashed line}). Bottom panels: profile of the total oxygen abundance.
   The origin of abscissae corresponds to the peak value of \te(\nii). }
   \label{rend}
\end{figure*}

\subsection{The trapped ionization front}

Most HH objects are located in molecular and neutral environments and their gas is shock-excited. However, 
\hhd\ --like other objects such as HH~444 \citep[see][]{Reipurth1998}, HH~505 \citep[see][]{Bally2001}, 
HH~529 \citep[see][]{Blagrave2006} or HH~202 \citep[see][]{Mesa-Delgado2009a}-- is an example of collimated 
HH jet immersed in a photoionized region, where the excitation of the gas is dominated by the stellar ionizing radiation. 
There are rather few theoretical models that describe the structure and evolution of photoionized HH jets 
\citep[e.g.][]{Raga2000,Masciadri2001,Raga2004,Raga2010}. Our unprecedentedly detailed integral field spectroscopical observations permit to study not only 
the ionization structure but also the physical conditions of the gas along the head of the bow shock, 
something rather unusual in the literature. Therefore, our data are of paramount interest to compare with the predictions 
of jet models. The most suitable models for the particular conditions of \hhd\
 are those by \cite{Masciadri2001}, who consider a 
source of radiation with a flux of ionizing photon comparable to that expected for \oric\ and located at different distances from the jet. 
Especially interesting is their model 2, 
in which the linear distance between the ionizing source and the jet is of the order of that of \hhd\ 
with respect to \oric. An apparent drawback of our comparison with the models of \cite{Masciadri2001} is that they consider that the direction 
toward the ionizing star is perpendicular to the outflow axis, 
whilst in our case the jet is oriented very close to the direction of \oric. In any case, this difference does not affect the results qualitatively. 

\cite{Masciadri2001} find that their model 2 produces a trapped ionization front at the head of the bow shock while the wings are completely ionized. This is 
precisely what we observe in our Fig. \ref{fluxes204}a, where the \oi\ 6300 \AA\ emission is absent in the wings but strong at the head. 
The \oi\ emission is concentrated to the south east side of the head, roughly directed away from \oric. 
Evidences of a trapped ionization front were also obtained in the case of 
HH~202 \citep{Mesa-Delgado2009a,Mesa-Delgado2009b}, located at the north west of \oric. In HH~202, the stronger \oi\ 6300 \AA\ emission 
is also located in the direction away from the Trapezium stars. 

 Due to the fact of having a trapped ionization front and following \cite{Mesa-Delgado2009b}, we can estimate the width of the ionized slab 
of \hhd\ and its physical separation with respect to the main ionization source, \oric. In our case, the ionized slab was computed from the 
maximum emission profiles of \Oo\ and \Hep\ with an angular distance on the plane of sky of about $2.9\pm0.5$ arcsec.  
Using the distance to the Orion Nebula obtained by \cite{Menten2007}, $d=414\pm7$ pc, and the inclination angle of \hhd\
with respect to the plane of the sky calculated by \cite{O'Dell2008a}, $\Theta=26^o$, we estimate $(6.1\pm1.2)\times10^{-3}$ pc 
for the width of the ionized slab. 
To trap the ionization front the incident Lyman continuum flux must be balanced by the recombination in the ionized slab, resulting in a physical
separation of $D=0.37\pm0.10$ pc for the \nel(\sii) of the head (see Table \ref{zones}).
This result suggests that \hhd\ is embedded within the body of the Orion Nebula and, therefore, discards
the origin of the ionized front as result of the interaction of the gas flow with the veil, which is between 1 and 3 pc \citep[][]{Abel2004}. 
If we consider the same ion (from \Oo\ to \Op) to compute the ionized slab ($(2.9\pm1.2)\times10^{-3}$ pc), the physical separation is about $D=0.53\pm0.17$.

\subsection{The high-temperature arc at \hhd}\label{high_t_arc}

In the models of photoionized HH jets of \cite{Raga2004}, the temperature along the HH object is the typical of a gas in 
photoionization equilibrium $-$about $10^4$ K$-$ but larger at the leading working surface of the bow shock 
because of shock heating. 
This high-temperature zone is narrow and precedes the area of high-density shocked gas behind the working surface. The high-temperature arc we see 
in Fig. \ref{condfis204}a may very likely correspond to that predicted structure. In fact, the arc delineates the outer edge of the bow shock and precedes the 
high-density compressed zone that forms the head of \hhd\ as we see in Fig. \ref{condfis204}. The upper limit for the width of the arc is 1$\arcsec$, which corresponds to 
a linear dimension of $\sim$6$\times$10$^{15}$ cm at the distance of the Orion Nebula. The models of \cite{Masciadri2001} $-$more appropriate for the Orion Nebula 
conditions$-$ predict a high-temperature zone at the leading working surface with a width of the order of $\sim$4$\times$10$^{15}$ cm, in a good 
agreement with our result. 

There is a previous example of a similar structure in the literature. Based on spatially resolved echelle spectroscopy, \cite{Solf1988}
reported a steep peak of the \te({[C~{\sc i}\relax]) just at the leading edge of the bow shock of the prototypical HH~1, showing values as 
high as 2 $\times$ 10$^4$ K in a region about 1$\arcsec$ wide. Because HH~1 belongs to the star-forming region Lynds 1641 $-$located at the Orion Nebula Cluster$-$ the linear dimensions of this high-temperature zone is rather similar to that we estimate for \hhd. \cite{Solf1988} finds that the spatial changes of electron temperature are far much less pronounced for other ions, specially in the case of \te(\nii), which shows hardly any variation. An important difference between HH~1 and \hhd\ is that the gas excitation mechanism in both objects are completely different, pure shock excitation in the case of HH~1 and photoionization in the case of \hhd. 

The high-temperature arc at the leading edge of \hhd\ is a real feature and not an artifact because it can also be seen in the long-slit data of \cite{Mesa-Delgado2008}. 
In Figs. \ref{rend}a and b we show the spatial profiles of \nel(\sii) and \te(\nii) (Fig. \ref{rend}a) and the O abundance (Fig. \ref{rend}b) of 
a small section along the slit position 3 of \cite{Mesa-Delgado2008} crossing the head of \hhd. That slit passes through approximately the middle of the 
high-temperature arc. In Fig. \ref{rend}a, we can see that the \te(\nii) peak is $2.5\arcsec$ ahead the \nel(\sii) maximum and that the \te(\nii) peak 
coincides with a minimum of the total O abundance, just the effect observed in our PMAS data (see Fig. \ref{abund204}c). For comparison, in Figs. \ref{rend}c 
and d we show the spatial profiles of \nel(\sii), \te(\nii) and O abundance obtained from a pseudo-slit --$1\arcsec$ wide-- obtained along the main diagonal of 
our PMAS maps, which is rather close to slit position 3 of \cite{Mesa-Delgado2008}. The similar behaviour of both sets of profiles is remarkable. 

\section{Conclusions}\label{conclusions}

In this paper we analyse integral field optical spectroscopy of the area encompassing the 
head of the bow shock of the conspicuous Herbig-Haro object of the Orion Nebula: \hhd.
We have mapped emission line fluxes and ratios, physical conditions, and ionic and total O abundances at 
spatial scales of 1$\arcsec$. 
 
The maps show a clear ionization stratification across the object. The presence of  
strong  [\ion{O}{i}] emission at the head of the bow shock and its absence at the wings indicate 
that those zones are optically thick and have developed a ionization front.
 We find evidences that \hhd\ is formed within the ionized gas of the nebula and not the Veil.
The head of the bow shock shows electron densities at least five times larger than the background 
ionized gas. We find that the density indicator based on [\ion{S}{ii}] is not providing the correct 
\nel([\ion{S}{ii}]) values because the nebular and auroral lines of [\ion{S}{ii}] give different ionic 
abundances. This is because the densities at the head are larger than the critical density. We recompute the 
\nel([\ion{S}{ii}]) map solving this inconsistency and use it in the subsequent calculations. 

We discover a narrow arc of high \te([\ion{N}{ii}]) values just at the southeast edge of the head. The temperature in this zone 
reaches values about 1,000 K higher than in the rest of the object. This arc may correspond to  
structures predicted by the models of photoionized HH jets of \cite{Raga2004} or \cite{Masciadri2001}, which are  
produced by shock heating at the leading working surface of the gas flow. The \Op\ and O abundances are somewhat 
underestimated --between 0.1 and 0.2 dex-- at this arc due to the higher temperature. We find also that 
the  \Op\ and O abundances are both about 0.1 dex larger at the northern half of the head of \hhd. Following 
\cite{Mesa-Delgado2011}, the most likely explanation is that the appropriate density for 
the \Op\ ion should be about 1000 \cmc\ lower than the nominal value of \nel(\sii) assumed for those 
spaxels. The use of a wrong higher density produces an overestimate of the collisional de-excitation of  the \oii\ $3727$ \AA\ line. 

The results of this paper demonstrate that the compression and heating of the gas due to high-velocity flows affect 
the chemical abundance determinations in \hii\ regions. When densities reach values of the order of 10$^3$$-$10$^4$ \cmc\ --as in the compressed heads of HH objects or at ionization 
fronts \citep[see][]{Mesa-Delgado2011}-- the \Op\ and O abundances determined from the intensity of collisional excited lines may be overestimated. Also, the presence of localized 
shock heating may affect the thermal balance of the gas producing artificial lower abundances. Fortunately, these disturbing effects can be analyzed in some detail in the Orion Nebula because of the higher spatial resolution attainable for this object due to its closeness, but it is difficult to explore in more distant objects and virtually impossible in extragalactic 
\hii\ regions. Similar studies of larger areas of the Orion Nebula are necessary in order to: a) quantify how these localized effects can affect integrated abundance determinations when spatially resolved observations are not possible 
in distant objects, and b) prove whether these features could be the origin $-$or mimic the effects$-$ of the so-called temperature 
fluctuations \citep{peimbert1967}. Finally, it must be remarked that abundance determinations based on recombination line ratios $-$which 
are almost independent the adopted \nel\ and \te$-$ are virtually unaffected by these disturbing effects.

\section*{Acknowledgments} 
We thank A. Riera and A. Raga for several suggestions and discussions. We are grateful to the referee,  C. R. O'Dell, 
for the careful reading and suggestions  that have improved the quality of the paper.
This work has been funded by the Ministerio de Educaci\'on y Ciencia (MEC) under project AYA2007-63030.



\label{lastpage}

\end{document}